\documentclass{article}
\usepackage{amsmath}

\input{tcilatex}
\begin{document}

\title{New Anisotropic Sudden Singularities and Dimensional Reduction}
\author{John D. Barrow \\
%EndAName
DAMTP, Centre for Mathematical Sciences,\\
University of Cambridge,\\
Wilberforce Road, Cambridge CB3 0WA\\
United Kingdom\ \ \ \ \ \ \ \ }
\maketitle
\date{}

\begin{abstract}
We demonstrate the existence of sudden, finite-time singularities, with
constant scale factor, expansion rate and density, in expanding Bianchi type
IX universes with free anisotropic pressures. A new type of non-simultaneous
anisotropic sudden singularity arises because of the divergences of the
pressures, which may be of barrel or pancake type. The effect of one or more
directions of expansion hitting a sudden singularity is tantamount to
dimensional reductions as the non-singular directions continue expanding and
can see the sudden singularity in their past.
\end{abstract}

\section{Introduction}

Sudden cosmological singularities, first introduced in ref. \cite{BGT} and
developed systematically in refs.\cite{JB1, JB2, JB3,cot1, cot2, q, sudBD},
and reviewed in refs.\cite{rev1}, have attracted widespread interest. They
appear in a wide range of gravity theories and their solutions. They
typically occur when the pressure and scale factor acceleration diverge at a
finite time, $t_{s},$ while the scale factor, density and Hubble expansion
rate remain finite. Thus all terms in the Friedman equation, or its
equivalent in other theories of gravity, remain finite whilst finite-time
singularities occur in the acceleration and conservation equations.

Sudden singularities and their generalised counterparts \cite{JB2} are weak
singularities in the senses of Tipler \cite{T} and Krolak \cite{K} and their
conformal diagrams have been constructed in ref. \cite{dabconf}. Geodesics
are unscathed by sudden singularities, \cite{las}, and the general behaviour
of the Einstein and geodesic equations in their neighbourhood was found in
refs. \cite{cot1, cot2, lip}. This behaviour appears robust in the presence
of quantum particle production \cite{q}. The first examples were existence
proofs that required unmotivated pressure-density relations so that the
density could remain finite while the pressure diverged. However, more
recently, generalised singularities of this sort have been found by Barrow
and Graham \cite{BGrah} and appear in simple isotropic Friedman universes
with a scalar field having power-law self-interaction potentials for a
scalar field,$\phi $, of the form: $V(\phi )=V_{0}\phi ^{n}$, $0<n<1.\ $\
They always develop a finite-time singularity where the Hubble rate and its
first derivative are finite, but its second derivative diverges. For
non-integer $n>1,$ there is a class of models with even weaker
singularities. Infinities first occur at a finite time in the $(k+2)^{th}$
time-derivative of the Hubble expansion rate, where $k<n<k+1$ and $k$ is a
positive integer \cite{BGrah}. These models inflate but inflation ends in a
singular fashion.

\ In this paper we study a new effect in anisotropic cosmological models
experiencing non-simultaneous sudden singularities in all, or some, of their
directional scale factors. This can create a form of dimensional reduction
in which some directional scale factors experience sudden singularities,
while others do not. Those that do not experience the singularities continue
expanding as if in a lower-dimensional universe. We use the Bianchi type IX,
'Mixmaster', universe expanding away from the initial strong curvature at $%
t=0$ to illustrate this point and derive the general forms of the evolution
of the three expansion scale factors.

In what follows we set $c=1=8\pi G.$ Planck's constant does not appear and
0ur study is entirely classical. Quantum features can be studied using our
paper ref. \cite{q}.

\section{The Mixmaster model equations}

The spatially homogeneous diagonal Bianchi IX metric is \cite{LL},

\begin{equation}
ds^{2}=dt^{2}-\gamma _{ab}(t)e_{\mu }^{a}e_{\nu }^{b}dx^{\mu }dx^{\nu },
\label{met}
\end{equation}

where

\begin{equation}
\gamma _{ab}(t)=diag[a^{2}(t),b^{2}(t),c^{2}(t)],  \label{met2}
\end{equation}

and

\begin{equation}
e_{\mu }^{a}=%
\begin{pmatrix}
\cos z & \sin z\sin x & 0 \\ 
-\sin z & \cos z\sin x & 0 \\ 
0 & \cos x & 1%
\end{pmatrix}%
.  \label{met3}
\end{equation}

The general relativistic field equations in vacuum Bianchi type IX with
scale factors $a(t),b(t),c(t)$ and matter with density $\rho $ and
anisotropic pressures $p_{1},p_{2}$ and $p_{3}$, \cite{LL, bel2, mis}, are
most simply expressed by introducing the $\tau $ time defined in terms of
the comoving proper time, $t$, by

\begin{equation}
d\tau =\frac{dt}{abc},  \label{time}
\end{equation}

and $^{\prime }$ denotes $d/d\tau $. The field equations are

\begin{equation}
(\ln a^{2})^{\prime \prime }+a^{4}-(b^{2}-c^{2})^{2}=a^{2}b^{2}c^{2}(\rho
-p_{a}),  \label{field}
\end{equation}%
and their two cyclic permutations obtained under the transform $a\rightarrow
b\rightarrow c\rightarrow a,$ together with $p_{a}\rightarrow
p_{b}\rightarrow p_{c}$. A first integral, the Mixmaster Friedman-like
equation, exists and is: 
\begin{equation}
4\left( \frac{a^{\prime }b^{\prime }}{ab}+\frac{b^{\prime }c^{\prime }}{bc}+%
\frac{a^{\prime }c^{\prime }}{ac}\right)
=a^{4}+b^{4}+c^{4}-2a^{2}b^{2}-2c^{2}(b^{2}+a^{2})+4a^{2}b^{2}c^{2}\rho .
\label{fried}
\end{equation}

We see that when $a=b=c$ it reduces to

\begin{equation*}
12\frac{a^{\prime 2}}{a^{2}}=4a^{6}\rho -3a^{4}.
\end{equation*}%
Restoring the cosmic time derivative $a^{3}da/dt=da/d\tau ,$ we have
(overdot denotes $d/dt$) the standard closed isotropic universe's Friedman
equation after the coordinate transform $a\rightarrow \frac{\ a}{2}:$

\begin{equation*}
3\frac{\dot{a}^{2}}{a^{2}}=\frac{\rho }{3}-\frac{1}{a^{2}}.
\end{equation*}

When all the quartic terms are dropped in eqs. (\ref{field}) then they are
just like the Bianchi I equations with a curvature term that will dominate
the matter at late times so long as $p_{i}>-\rho /3$. We looked at the flat
Bianchi $I$ and $VII_{0}$ models in ref.(\cite{JB3}).

\section{The sudden singularity scale factor evolutions}

In order to establish the existence of anisotropic sudden singularities in
the Mixmaster metric at late time we look for the following forms for the
scale factors on approach to a finite-time singularity s $t\rightarrow t_{s}$
from below. The scale factors, their first time derivatives and the density, 
$\rho $, will be assumed to be finite at $t_{s}$, but second derivatives\ of
the scale factors, first derivatives of the density, and the principal
pressures will be allowed to diverge. Thus, we assume that the asymptotic
forms of the scale factors as $t\rightarrow t_{s}$ have the form that we
know is part of the general solution of the Einstein equations \cite{cot1,
cot2}, So, all terms in eq. (\ref{fried}) will be finite and eqs.(\ref{field}%
) reduce asymptotically to:

\begin{equation}
(\ln a^{2})^{\prime \prime }=-a^{2}b^{2}c^{2}p_{a}=-C_{a}p_{a},
\label{asymfield}
\end{equation}

et cycl. In this $t\rightarrow t_{s}$ limit we also have

\begin{equation}
a^{\prime \prime }\rightarrow \ddot{a}a^{2}b^{2}c^{2},  \label{asymfield1}
\end{equation}

et cycl., and so we have the simple system of asymptotic equations:

\begin{equation}
(\ln a^{2}\ddot{)}\rightarrow -p_{a}.  \label{asymfield2}
\end{equation}

Therefore, explicitly in the limit $t\rightarrow t_{s}$ we have,

\begin{equation}
(\frac{\ddot{a}}{a},\frac{\ddot{b}}{b},\frac{\ddot{c}}{c})\rightarrow -\frac{%
1}{2}(p_{a},p_{b},p_{c}).  \label{asymfield3}
\end{equation}

However, we want to allow the sudden singularity to arise at different times
for the motions in the directions of the different scale factors, so we
introduce three sudden singularity times, $t_{sa},t_{sb}$ and $t_{sc}>0$,
thus: \footnote{%
Technically, we can create a slightly more general but rather cumbersome
form by including powers of logarithms. Thus, for $a(t),$ we would have
\par
$a(t)=\left( \frac{t}{t_{sa}}\right) ^{q_{a}}(a_{sa}-1)+1$%
\par
$-\left( 1-\frac{t}{t_{sa}}\right) ^{n_{a}}\left\{
\dsum\limits_{j=0}^{\infty
}\dsum\limits_{k=0}^{N_{j}}a_{jk}(t_{s}-t)^{j/Q}\left( \log
^{k}[t_{s}-t]\right) \right\} $%
\par
where $N_{j}\leq j$ is a positive integer and $Q$ is a positive rational.
For the corresponding expressions for $b(t)$ and $c(t)$ we replace $a_{jk}$,$%
N_{j},$ and $Q$ by different independent constants satisfying the same
inequalities.}%
\begin{eqnarray}
a(t) &=&\left( \frac{t}{t_{sa}}\right) ^{q_{a}}(a_{sa}-1)+1-\left( 1-\frac{t%
}{t_{sa}}\right) ^{n_{a}},  \label{asymformc} \\
b(t) &=&\left( \frac{t}{t_{sb}}\right) ^{q_{b}}(a_{sb}-1)+1-\left( 1-\frac{t%
}{t_{sb}}\right) ^{n_{b}},  \label{asymformb} \\
c(t) &=&\left( \frac{t}{t_{sc}}\right) ^{q_{c}}(a_{sc}-1)+1-\left( 1-\frac{t%
}{t_{sc}}\right) ^{n_{c}}.
\end{eqnarray}

Here, the constants $0<q_{a},q_{b},q_{c}<1$ and $1<n_{a},n_{b},n_{c}<2.$
Therefore,

\begin{equation}
\ddot{a}=\frac{q_{a}(q_{a}-1)}{t_{sa}^{2}}\left( \frac{t}{t_{sa}}\right)
^{q_{a}-2}-n_{a}(n_{a}-1)(t_{sa}-t)^{n_{a}-2}  \label{accn}
\end{equation}%
and the forms for $\ddot{b}(t)$ and $\ddot{c}(t)$ are given by in the same
form after substitution of the $q^{\prime }s$ and $n^{\prime }s$. The values
of $\ddot{a},\ddot{b}(t),$ and $\ddot{c}(t)$ can each diverge if the values
of $n_{a},n_{b},n_{c}$ are less than two and greater \ than one, as assumed.
This will create complementary divergences in the values of the pressures
because of eq. (\ref{asymfield3}). When $n_{a}=n_{b}=n_{c}$, this is similar
to the behaviour in the Friedman model, first described in (\cite{JB2}).
However, it is possible to create new types of anisotropic sudden
singularity in acceleration and associated principal pressure by making $%
n_{a}\neq n_{b}\neq n_{c}$, or allow some directions to avoid a sudden
singularity while others experience it. The possibilities are:

1. $1<n_{a},n_{b},n_{c}<2:$ sudden singularities in all directions and their
associated principal pressures at different times if $t_{sa},t_{sb}$ and $%
t_{sc}$ are unequal, simultaneously if they are equal.

2. $1<n_{a},n_{b}<2$ and $n_{c}>2$ and similar for the other two permutations%
$:$ sudden singularities in two directions and their associated principal
pressures (but not in the third).

3. $1<n_{a}<2$ and $n_{b},n_{c}>2$ and similar for the other two
permutations: sudden singularity in one direction and its principal pressure
(but not in the third).

For example, in the second case (2), we have a non-simultaneous sudden
singularity in the $a$ and $b$ directions, with

\begin{eqnarray}
a(t) &=&\left( \frac{t}{t_{sa}}\right) ^{q_{a}}(a_{sa}-1)+1\rightarrow
a_{sa},  \label{nonsim1} \\
b(t) &=&\left( \frac{t}{t_{sa}}\right) ^{qb}(a_{sb}-1)+1\rightarrow a_{sb},
\label{nonsim2}
\end{eqnarray}

while the expansion parallel to the $c$ direction continues with $n_{c}>2,$
and there is no singularity in $c(t)$ as $t\rightarrow t_{sc}$ and so it
continues past the singularities affecting particles moving parallel to the $%
a$ and $b$ directions for $t>t_{sc}$,

\begin{equation}
c(t)=\left( \frac{t}{t_{sc}}\right) ^{q_{c}}(a_{sc}-1)+1-\left( 1-\frac{t}{%
t_{sc}}\right) ^{n_{c}}\rightarrow \left( \frac{t}{t_{sc}}\right)
^{q_{c}}(a_{sc}-1)+1-\left( 1-\frac{t}{t_{sc}}\right) ^{n_{c}}\   \label{c}
\end{equation}

The relative values of $q_{c}$ and $n_{c}$ determine this evolution.
Typically we expect $n_{c}>q_{c}$ and so

\begin{equation}
c(t)\rightarrow \left( \frac{t}{t_{sc}}\right) ^{q_{c}}(a_{sc}-1)+1-\left( 1-%
\frac{t}{t_{sc}}\right) ^{n_{c}}\rightarrow 1-\left( \frac{t}{t_{sc}}\right)
^{n_{c}}(-1)^{n_{c}}  \label{c1}
\end{equation}%
For example, for odd $n_{c}$, we have for $t>t_{sc}$,

\begin{equation}
c(t)\rightarrow \left( \frac{t}{t_{sc}}\right) ^{n_{c}}\ .  \label{c2}
\end{equation}%
For even $n_{c}$, we have for $t>t_{sc}$, from (\ref{c}),

\begin{equation}
c(t)\rightarrow \ \left( \frac{t}{t_{sc}}\right) ^{q_{c}}(a_{sc}-1)-1+\left( 
\frac{t}{t_{sc}}\right) ^{n_{c}},  \label{c3}
\end{equation}

and the possibility of a switch between the two asymptotic time-dependencies.

\section{Dimensional reduction and its physical interpretation}

The possibility of sudden singularities occurring anisotropically at
different times is a new feature of this phenomenon. There are no strong
curvature singularities associated with any of the sudden singularities and
we expect geodesics to be unscathed by the experience unless the underlying
expansion anisotropy contributed strong tidal forces \cite{las}. The
appearance of finite-time singularities for the motion of only some of the
expansion scale factors is created by the anisotropic pressures. It means
that particles moving in the singular directions will hit the pancake or
barrel-like sudden singularity, leaving those moving in the directions
orthogonal to them unscathed. This has an interesting consequence. Suppose
that we repeated our calculations for anisotropic cosmological models with
many space dimensions, $N$, then we might have $S$ of those dimensions
experiencing sudden singularities (not necessarily all at the same time),
eventually leaving $N-S$ to continue expanding. In effect, this is a
cosmological dimensional reduction process. If $N-S=3$ then we would be left
with a 3-dimensional expanding space. Observers in that space could in
principle look back down their past lightcones and see consequences of the
sudden singularities in the other non-evolving dimensions. This will have
consequences for the constants of Nature. If the true constants are defined
in the $N$-dimensional space then in all subspaces of lesser dimension, the
apparent constants in their space will be seen to evolve in time on the same
timescale that the extra dimensions change on. Observers in an $(N-S)$%
-dimensional expanding subspace will at first see small variations in
quantities like their local fine structure 'constant', or the Newtonian
gravitation 'constant', following the overall volume expansion. But when the
extra dimensions hit their finite-time singularities there could be dramatic
evolution of the local $3$-dimensional constants \cite{const1, BD, marc,
uzan}. However, the sudden singularities are characterised by the scale
factors tending to constant values at the singularity. Since the evolution
of the local 'constants' is determined by inverse powers of the mean scale
of the extra dimensions, there will no dramatic evolution of the values of
local constants towards zero or infinity \cite{const2, const3}. The global
structure of singular cosmological models of this type may prove to be
interesting and quite different to that accompanying strong curvature
singularities. \ We don't expect quantum particle production effects, or
their classical analogue of bulk viscosity, because the local expansion
rates on which these processes depend are assumed constant on approach to
the sudden singularity \cite{q}. Higher-order versions of sudden
singularities \cite{JB2, BGrah} will also be possible in these cosmological
models.

\section{Conclusions}

We have studied the presence of sudden singularities, with finite scale
factors, expansion rates and matter densities in the most general closed
spatially homogeneous universes of Bianchi type IX. They permit divergences
in scale factor accelerations an pressures at finite-time singularities,
where no curvature invariants diverge. In the presence of anisotropic
pressures we found a new variety of non-simultaneous directional sudden
singularity which can occur in all or any of the expanding directions. This
allows expansion in some directions to end at a sudden singularity while
those in other non-singular directions do not. This creates a new form of
dimensional reduction driven by the anisotropic pressures, some of which may
diverge at finite time while others remain finite. The expansion continues
unaffected in the non-singular directions and the sudden singularities and
their consequences could be observed in the past of the non-singular
directions.

\textit{Acknowledgement.} The author is supported by the Science and
Technology Facilities Council of the UK (STFC)


\begin{thebibliography}{99}
\bibitem{BGT} J.D. Barrow, Galloway G. and Tipler F.J., Mon. Not. R. Astron.
Soc\textit{. }\textbf{223,} 835 (1986)

\bibitem{JB1} J.D. Barrow, Class. Quantum Gravity\textit{\ }\textbf{21,} L79
(2004)

\bibitem{JB2} J.D. Barrow, Class. Quantum Gravity\textit{\ }\textbf{21,}
5619 (2004)

\bibitem{JB3} J.D. Barrow and C.G. Tsagas, Class. Quantum Gravity\textit{\ }%
\textbf{22,}1563 (2005)

\bibitem{cot1} J.D. Barrow, S.T. Cotsakis and A.Tsokaros, Class. Quantum
Gravity \textbf{27,}165017 (2010)

\bibitem{cot2} J.D. Barrow, and S.T. Cotsakis, Phys. Rev. D \textbf{88,}
067301 (2013)\textit{\ }

\bibitem{q} J.D. Barrow, A.B. Batista, J.C. Fabris and S. Houndjo, Phys.
Rev. D\textit{\ }\textbf{78,}123508 (2008)

\bibitem{sudBD} J.D. Barrow, Classical and Quantum Gravity \textbf{37},
065014 (2020)

\bibitem{rev1} S. Nojiri and S.D. Odintsov, Phys. Rev. D \textbf{78,} 046006
(2008)

\bibitem{T} F.J. Tipler, Phys. Lett. A \textbf{64,} 8 (1977)

\bibitem{K} A. Krolak, Class. Quantum Gravity\textit{\ }\textbf{3,} 267
(1986)

\bibitem{dabconf} M.P. D\c{a}browski and K. Marosek, arXiv:1806.00601

\bibitem{las} L. Fernandez-Jambrina and R. Lazkoz, Phys. Rev. D \textbf{70,}%
121503(R) (2004)

\bibitem{lip} J.D. Barrow and S.Z.W. Lip, Phys. Rev. D \textbf{80,} 043518
(2009)

\bibitem{BGrah} J.D. Barrow, and A.A.H. Graham, Phys. Rev. D \textbf{91,}
083513 (2015)

\bibitem{LL} L. Landau and E.M. Lifshitz, \textit{The Classical Theory of
Fields}, Pergamon Press, Oxford, 4th rev. edn., (1974)

\bibitem{bel2} V.A. Belinski and M. Henneaux, \textit{The Cosmological
Singularity}, Cambridge Univ. Press, Cambridge, (2018)

\bibitem{mis} C.W. Misner, Phys. Rev. Lett.$\mathbf{\ }$\textbf{22,} 1071
(1969)

\bibitem{DLN} A.G. Doroshkevich, V.N. Lukash and I.D. Novikov, Sov. Phys.
JETP \textbf{37}, 739 (1973)

\bibitem{const1} J.D. Barrow, Phys. Rev. D \textbf{35}, 1805 (1987)

\bibitem{BD} C.H Brans and R.H. Dicke, \textit{\ }Phys. Rev. \textbf{124,}
925 (1961)

\bibitem{marc} W. Marciano, Phys. Rev. Lett.\textbf{\ 52}, 489 (1984)

\bibitem{uzan} J.-P. Uzan, Varying constants, gravitation and cosmology,
Living Reviews in Relativity, https://arxiv.org/abs/1009.5514

\bibitem{const2} H.B. Sandvik, J.D. Barrow and J. Magueijo, Phys. Rev. Lett. 
\textbf{88}, 031302 (2002)

\bibitem{const3} J.D. Barrow and J. Magueijo, Phys. Rev. D \textbf{72},
043521 (2005)
\end{thebibliography}
\end{document}